# Highly-Entangled Polyradical Nanographene with Coexisting Strong Correlation and Topological Frustration

Shaotang Song[1,7], Andrés Pinar Solé[2,4,7], Adam Matěj[2,4,7], Guangwu Li[1,3,7], Oleksandr Stetsovych[2], Diego Soler[2], Huimin Yang[1], Mykola Telychko[1], Jing Li[1], Manish Kumar[2], Jiri Brabec[6], Libor Veis*[6], Jishan Wu*[1], Pavel Jelinek*[2,4], Jiong Lu*[1,5]

[1]Department of Chemistry, National University of Singapore, 3 Science Drive 3, Singapore 117543, Singapore.

[2]Institute of Physics of the Czech Academy of Sciences, Prague, 16200, Czech Republic.

[3]Center of Single-Molecule Sciences, Frontiers Science Center for New Organic Matter, College of Electronic Information and Optical Engineering, Nankai University, Jinnan District, Tianjin 300350, China.

[4]Regional Centre of Advanced Technologies and Materials, Czech Advanced Technology and Research Institute (CATRIN), Palacký University Olomouc, 78371 Olomouc, Czech Republic.

[5]Institute for Functional Intelligent Materials, National University of Singapore, 117544, Singapore.

[6]Department of Theoretical Chemistry, J. Heyrovsky Institute of Physical Chemistry, Czech Academy of Sciences, Prague 18200, Czech Republic.

[7]These authors contributed equally: Shaotang Song, Andrés Pinar Solé, Diego Soler, Guangwu Li.

E-mail: libor.veis@jh-inst.cas.cz; chmwuj@nus.edu.sg; jelinekp@fzu.cz; chmluj@nus.edu.sg

**Abstract**

Open-shell benzenoid polycyclic aromatic hydrocarbons, known as magnetic nanographenes, exhibit unconventional π-magnetism arising from topological frustration or strong electronic-electron (e-e) interaction. Imprinting multiple strongly entangled spins into polyradical nanographenes creates a major paradigm shift in realizing non-trivial collective quantum behaviors and exotic quantum phases in organic quantum materials. However, conventional design approaches are limited by a single magnetic origin, which can restrict the number of correlated spins or the type of magnetic ordering in open-shell nanographenes. Here, we present a novel design strategy combing topological frustration and e-e interactions to fabricate the largest fully-fused open-shell nanographene reported to date, a 'butterfly'-shaped tetraradical on Au(111). We employed bond-resolved scanning tunneling microscopy and spin excitation spectroscopy to unambiguously resolve the molecular backbone and reveal the strongly correlated open-shell character, respectively. This nanographene contains four unpaired electrons with both ferromagnetic and anti-ferromagnetic interactions, harboring a many-body singlet ground state and strong multi-spin entanglement, which can be well described by many-body calculations. Furthermore, we demonstrate that the nickelocene magnetic probe can sense highly-correlated spin states in nanographene. The ability to imprint and characterize many-body strongly correlated spins in polyradical nanographenes not only presents exciting opportunities for realizing non-trivial quantum magnetism and phases in organic materials but also paves the way toward high-density ultrafast spintronic devices and quantum information technologies.

**Introduction**

The field of polycyclic aromatic hydrocarbons (PAHs) has witnessed significant advances in recent years, particularly in the area of open-shell nanographenes.[1-11] However, the design and synthesis of highly-correlated polyradical nanographenes with unconventional π-magnetism and strong many-body spin entanglement remain challenging for both in-solution and on-surface synthesis protocols. While kinetic protection and thermal dynamic stabilization have succeeded in the in-solution synthesis of polyradical nanographenes, substitutional groups may impede intrinsic properties and applications.[7,12-16] Furthermore, using traditional in-solution chemistry to synthesize large, unsubstituted fully-fused open-shell nanographenes that host intrinsic quantum magnetism and strong correlations remains a significant challenge due to poor solubility and high reactivity. On the other hand, advances in on-surface synthesis and scanning probe microscopy have allowed for the fabrication and sub-molecular resolution characterization of unsubstituted nanographenes with intrinsic open-shell characters that originate from either topological frustration or strong e-e interaction in the π-bond network, as observed in various magnetic nanographenes such as the triangulene series, Clar's goblet, rhombenes, *peri*-acenes, and more.[3-8,17-28]

One approach to imprint π-magnetism in open-shell nanographenes is to exploit the topological frustration of either one set of honeycomb bipartite lattices through a sublattice imbalance (defined as class I, e.g. triangulene)[5,6,8,9,17-22,29,30] or the topological frustration of two sets of honeycomb bipartite lattices (defined as class II, e.g. Clar's goblet).[4,31] Another alternative is to use the enhanced e-e interaction of π-electrons, which may introduce spin-symmetry breaking of frontier orbitals resulting from the interplay of electronic hybridization of localized frontier states and Coulomb repulsion between valence electrons (Rhombene nanographenes).[3,32] However, on-surface fabrication of highly-entangled fully-fused polyradical π-systems *via* these conventional design principles still remains scarce, as a single magnetic origin often limits the number of correlated spins or the type of magnetic ordering in open-shell nanographenes. Although it is, in principle, possible to design higher polyradicals by increasing their size through a single magnetic origin, the increased system size brings significant challenges in dealing with their stability and solubility, hindering their preparation.

Here, we introduce a new conceptual design of polyradical nanographene that features multiple strongly entangled many-body quantum spins arising from the interplay of strong e-e correlation and topological frustration (Fig. 1). Our design involves fusing four [3]triangulene motifs onto the edges of a [3]rhombene to create a butterfly-shaped nanographene with a carefully crafted geometry that renders both sublattices (A and B) topologically frustrated. According to the hexagonal graphs theorem,[33] the number of zero-energy eigenstates in the tight binding model for graphene nanoflakes can be determined by the term of ''nullity'' ($\eta$), which is equal to the difference between the maximum numbers of nonadjacent vertices ($\alpha$) and edges ($\beta$), as depicted in Fig. 1b.[31,] The sum of $\alpha$ and $\beta$ equals to the total number of carbon atoms (N = 118) in this large nanographene. The topologically-frustrated butterfly geometry in our design yields a nullity of two ($\eta = \alpha - \beta = 2$), which predicts the presence of two radicals. Additionally, this design creates a sufficiently large size (the largest size of fully fused open-shell nanographene by far) to trigger spin-symmetry breaking of frontier occupied orbitals through strong e-e interaction that dominates over the hybridization energy, yielding two more radicals. The combination of these two magnetic origins results in a designer tetraradical nanographene with both ferromagnetic and antiferromagnetic coupling of correlated spins that is rarely found in fully aromatic PAHs possessing only a single magnetic origin.

To gain further insight into the electronic structure of this system, it is also instructive to analyze the energy spectrum of a single-electron tight-binding model with one $p_z$ orbital per carbon atom in the first neighbor hopping approximation described by the Hückel Hamiltonian.[34,35] Fig. 1c reveals the presence of two zero-energy singly occupied molecular orbitals (SOMO) denoted as $\varphi_2$ and $\varphi_3$, which are strongly localized on the opposite zigzag edges. These two SOMOs originate from the topological frustration of the bipartite honeycomb lattice that readily contributes to two unpaired electrons. Additionally, there are two frontier orbitals $\varphi_1$ and $\varphi_4$ separated by a small energy gap (~ 0.5 eV). Therefore, according to a simple theorem,[36] they undergo spin symmetry breaking due to the enhanced e-e correlation between the two orbitals (refer to section 3.2 in SI for a more detailed discussion). Thus, in principle, the system may host four unpaired electrons, as it will be demonstrated rigorously both by experimental and theoretical findings described below.

**On-surface synthesis and characterization of molecule 1**

The synthesis of the butterfly-shaped nanographene on Au(111) relies on thermal annealing of precursor **1′** (Fig. 2a), which consists of a rhombene core and four 9-(2,6-dimethylphenyl) anthryl motifs attached to the central carbon at four zigzag edges. Upon the oxidative cyclodehydrogenation, the periphery motifs of this molecule are expected to transform into triangulenes and fuse with the rhombene core, leading to the formation of desired butterfly-shaped molecule **1** with an open-shell tetraradical character. Such a large π-conjugated precursor can be obtained through a multiple-step organic synthesis (refer to supplementary information). Firstly, the treatment of tetra-aldehyde compound **3** with 9-(2,6-dimethylphenyl) anthryl lithium affords a tetra-hydroxyl intermediate, which then undergoes Friedel–Crafts cyclization mediated by $BF_3 \cdot Et_2O$ to yield **2**. Subsequently, oxidative dehydrogenation by *p*-chloranil in toluene produces the desired precursor **1′**, which is then sublimated onto Au(111) and annealed at 600 K for the on-surface synthesis of molecule **1** *via* cyclodehydrogenation reactions. An overview STM image of the surface after the annealing reveals the formation of butterfly-shaped molecules and randomly connected oligomers (Fig. S2). Close-up examination of individual molecules by bond-resolved scanning tunneling microscopy (BRSTM) with a CO functionalized tip enables us to resolve the intact molecular backbone of the butterfly-shaped product **1** with four perfect triangulene corners (Fig. 2b).

We then carried out d$I$/d$V$ spectroscopic measurements to probe the local electronic structures of molecule **1**. The d$I$/d$V$ spectra taken over the corner and the edge of **1** reveal two prominent features located at energies of -0.60 eV and 0.75 eV, respectively (Fig. 2c). These features are primarily localized at the triangulene corners and at the gulf region of two zigzag edges of **1** (Fig. 2e). Notably, the occupied resonance feature exhibits considerably higher d$I$/d$V$ intensity in the bay region (labeled in Fig. 2a) than the empty resonance feature, which is corroborated by the simulated d$I$/d$V$ maps (Fig. 2f) using the concept of Dyson orbitals (see discussion later). Additionally, the d$I$/d$V$ spectra taken over the corner also show a small gap feature in the vicinity of the Fermi level ($E_F$). Amplification of the spectrum demonstrates a symmetric step-like feature around the $E_F$ (Fig. 2d), attributed to the spin-flip excitation from the ground to the excited state.[3,4,9,18,32,37-39] The assignment of this feature to the spin excitation process will be further supported by many-body theoretical analysis, as discussed in the following section.

**The tetraradical character with strong spin correlation in molecule 1**

As the single-determinant density functional theory (DFT) method is inadequate for accurately modeling polyradical systems with a strong multireference character, our DFT calculations consistently overestimate the singlet-triplet excitation energy by over one order of magnitude (see Table S1). To determine the ground state character of molecule **1** and confirm its tetraradical nature, we therefore, performed many-body calculations using the complete active space self-consistent field (CASSCF) method. Moreover, we employed the second-order N-electron valence perturbation theory in its domain-based local pair-natural orbital approximation (DLPNO-NEVPT2)[40] to account for the missing (out-of-CAS) electron correlation in the CASSCF calculations.

As discussed above, Hückel model, although simplistic, provide valuable insight into the electronic structure of molecule **1**. In particular, the model predicts the existence of two zero-energy SOMO orbitals and two additional orbitals (labeled as $\varphi_1$ and $\varphi_4$) separated by a narrow energy gap (Fig. 1c and Fig. S4). This small energy gap facilitates the spin symmetry-breaking transition due to e-e interaction. The CASSCF calculations in the active space consisting of the aforementioned four orbitals, CAS(4,4), confirm the open-shell nature of the singlet ground state ($S=0$) with four unpaired electrons (Fig. 1d and Table S2). The occupation number of each orbital is close to 1, which further verifies its tetraradical character (Fig. 3a). The calculated unpaired density[35] originating from the presence of the four unpaired electrons, as shown in the inset of Fig. 3b, are mainly localized on the four corners of the molecule.

In addition to the ground state ($S=0$), CASSCF calculations also reveal that the first excited state is a triplet state with a strong multireference character (Table. S3). Moreover, the high-level DLPNO-NEVPT2/CASSCF(4,4) calculations predict the triplet state ($S=1$) to be higher in energy by 8 meV than the ground state, in excellent agreement with the experimental value of ~9 meV. To further support the assignment of the IETS feature near $E_F$ to the spin-excitation process, CASSCF natural transition orbitals (NTOs) were calculated.[33,34] NTOs represent a compact orbital picture of excitation processes, and in this case, they represent the transition from the singlet ground state to the first excited triplet state. The simulated spin-excitation d$I$/d$V$ maps corresponding to a sum of contributions from individual singlet-triplet NTOs are displayed in Fig. 2h, showing good agreement with the experimental d$I$/d$V$ map acquired at 9 meV (Fig. 2g). This further supports the origin of the observed IETS features at this energy from the spin-excitation process.

To provide a better rationalization of the experimental d$I$/d$V$ maps, we calculated CASSCF Dyson orbitals, which describe the change in electron distribution during electron ionization/attachment of individual molecules.[41] Unlike canonical DFT orbitals, which are only suitable for describing weakly correlated systems,[42] Dyson orbitals represent a quantity that can be directly comparable to d$I$/d$V$ maps of frontier orbitals, even for strongly correlated molecules. This is further evidenced by a good agreement between the simulated constant-height d$I$/d$V$ maps computed from CASSCF Dyson orbitals for electron ionization and attachment (Fig. 2f and Fig. S3) and the experimental constant-current d$I$/d$V$ images (Fig. 2e). The singlet ground state of molecule **1** has an open-shell character with four highly-correlated unpaired electrons exhibiting a non-trivial arrangement of ferromagnetic and antiferromagnetic exchange coupling. Consequently, the wave function of the singlet ground state has a strong multi-reference character, represented by a superposition of multiple electronic configurations, wherein the dominant configuration has a weightage smaller than 33% (Fig. S5).

To determine the mutual coupling of individual unpaired spins in forming the singlet ground state of molecule **1**, we calculated the spin-spin correlation function ($A_{ij} = \langle \hat{S}_i \hat{S}_j \rangle - \langle \hat{S}_i \rangle \langle \hat{S}_j \rangle$) between each pair of spins *i* and *j*, using a maximally localized basis set representation of four unpaired electrons that are predominantly located on individual triangulene corners of molecule **1**. As shown in Fig. 3b, spins localized on the same sublattices exhibit ferromagnetic coupling, while those localized on different sublattices exhibit anti-ferromagnetic coupling.

**Probe the correlated π-spins and many-body spin excitation with a nickelocene tip**

We further employed the SPM technique with a nickelocene (NiCp$_2$) functionalized probe to directly verify the magnetic properties of **1** (Fig. 4a) by obtaining inelastic electron tunneling spectroscopy (IETS) spectra at different tip-sample distances.[43-46] The NiCp$_2$ molecule has a triplet ground state ($S = 1$, $m_S = 0$) and doubly degenerate triplet excited state ($S = 1$, $m_S = \pm 1$) due to the presence of a positive magnetic anisotropy parameter $D \sim 4$ meV defining easy magnetization plane along the z-axis with.[43] The spin-excitation from the ground state ($m_S = 0$) to one of the doubly degenerated excited states ($m_S = \pm 1$) of the NiCp$_2$ probe is manifested as the corresponding peak/dip features at ±4 mV symmetrically around $E_F$ in the IETS spectrum taken on a bare Au(111), as shown in Fig. 4b. It is noteworthy that, as the tip approaches the Au(111) surface, the energetic positions of the peak or dip features remain constant while their intensities increase. In contrast, the IETS spectra taken over the corner of **1** display additional feature at ±13 mV at a large tip-sample distance (Fig. 4c). As the tip-sample distance decreases, the IETS signal at ±4 mV moves towards $E_F$, while the new features at ±13 mV shift away from $E_F$, accompanied by a monotonic broadening of peak and dip features. The observed modification of the IETS spectra can be attributed to the magnetic exchange interaction between the spin states of the NiCp$_2$ probe and the tetraradical molecule, which provides direct experimental evidence of the presence of π-magnetism in product **1**. The combination of a NiCp$_2$ probe study with theoretical calculations allows us to unravel the exotic magnetic coupling with multiple-entangled spin interactions in product **1**, which will be discussed in more detail in the following context.

To gain a theoretical understanding of IETS spectra with a NiCp$_2$ probe, we have developed a Heisenberg spin model, which allows us to rationalize the experimental d*I*/d*V* spectra acquired over molecule **1** with a NiCp$_2$-fuctionalized tip. The model consists of an effective spin Hamiltonian describing the interaction between effective spin states residing on the

NiCp$_2$ probe and molecule **1**. The spin model of the molecule consists of four spin sites that are mutually coupled by both ferromagnetic and antiferromagnetic exchange interactions, as determined by the spin correlation model presented in Fig. 3b. In addition, the exchange interaction between the NiCp$_2$ probe and the molecule is governed by a *z*-distance dependent parameter $J(z)$, which increases as the tip-sample distance ($z$) decreases. We then use the exact diagonalization method to calculate the energy spectrum of the model Hamiltonian at each tip-sample distance (refer to section 5 in SI for a more detailed description of the spin model).

Fig. 4d shows the calculated IETS of the spin Hamiltonian $\hat{H}$ with an increase of exchange coupling ($J$) between NiCp$_2$-tip and molecule to simulate the experimental spectrum with a decrease of the tip-sample distance. The lowest energy state at zero energy corresponds to the ground state of both NiCp$_2$ and molecule **1**, which is absent in the spectrum. The two states (denote as A) centered at ±5 meV for $J = 0$ correspond to the ground state of molecule **1** and the degenerated excited state ($m_S = \pm 1$) of NiCp$_2$. The calculated excitation energy of this state shows a slight decrease with increasing $J$, which agrees well with the experimental evolution of IETS peak at 5 meV. This feature can only be reproduced by considering the strongly asymmetric coupling between NiCp$_2$ and one of the spin sites of molecule **1** (refer to Fig. S6). Two additional states, corresponding to the coupling between the excited triplet state of molecule **1** and the ground state of NiCp$_2$, are predicted to be located at ±9 meV (denoted as B in grey lines in Fig. 4d). However, these excitations were not observed experimentally, likely due to the extremely low elastic tunneling current measured with a nickelocene probe at the ground state, making the detection of the spin-excitation signal of molecule **1** infeasible.[43] Furthermore, another set of states located at 14 meV (denoted as C) was observed, which corresponds to the quadruple degenerated configuration where both molecule **1** and NiCp$_2$ are in the first spin excited state. The exchange coupling $J(z)$ acting between the tip and the molecule lifts their degeneracy, yielding four different states. The similarity in the energy of these states, combined with the limited energy resolution of our spectroscopic measurements, hindered resolving the individual state. Nevertheless, the evolution of these states with increasing $J(z)$ nicely fits to the *z*-dependent IETS spectra (Fig. 4c). The very good agreement between the experimental and theoretical spectra confirms that NiCp$_2$ probe is capable of detecting highly correlated molecular spin states. Moreover, the ability of this functionalized magnetic probe to detect excited many-body spin states upon spin-flip process underscores the potential of this nanographene for quantum information coding and detection.

**Conclusion**

We have demonstrated a novel approach to designing a polyradical nanographene that incorporates the concepts of both strong e-e interaction and topological frustration. Through advanced multireference quantum chemical calculations, we have revealed that the butterfly-shaped tetraradical molecule exhibits a strongly correlated many-body ground state featuring highly entangled four spins. We have also utilized the NiCp$_2$ functionalized STM tip to probe the π-magnetism and many-body spin excitation directly. Our findings reveal a new design strategy for imprinting and characterizing many-body strongly correlated spins in polyradical nanographenes. This approach not only provides a promising platform for exploring exciting non-trivial spin entanglement and exotic low-dimensional quantum phases in organic materials (e.g. quantum spin liquid phases), but also opens up new possibilities for high-density ultrafast spintronic[31] devices and quantum information technologies.

**Method**

**In-solution and on-surface synthesis.** The synthetic procedure of the precursor **1′** is detailed in Supplementary Information. Au(111) single crystal (MaTeck GmbH) was cleaned by multiple cycles of Ar$^+$ sputtering and annealing. Knudsen cell (MBE-Komponenten GmbH) was used for the deposition of precursor molecules onto clean Au(111) surfaces under ultrahigh vacuum conditions (base pressure, $<2 \times 10^{-10}$ mbar) for on-surface synthesis of product **1**. After the deposition of precursors, the sample was annealed at elevated temperatures as stated in the main text for 20 min to induce intramolecular dehydrogenation.

**STM and d$I$/d$V$ characterization.** The experiments were conducted in LT-STM/AFM systems operated under ultrahigh vacuum (base pressure, $P < 2 \times 10^{-11}$ mbar) at a temperature of T = 4.5 K (Scienta Omicron) and 1.3 K (JT-SPECS), respectively. The d$I$/d$V$ spectra were collected using a standard lock-in technique with a modulated frequency of 479 Hz. The modulation voltages for individual measurements are provided in the corresponding Fig. captions. The STM tip was calibrated spectroscopically against the surface state of Au(111) substrate. All the BRSTM images were taken in constant height mode ($V = 20$ mV) with a CO functionalized tip. The STM images were analyzed and processed with Gwyddion software.

The IETS ($d^2I/dV^2$) spectra taken with a nickelocene-functionalized tip were conducted using a SPECS LT-SPM at 1.3 K. To deposit nickelocene molecules onto the cold surface in the STM chamber, we pre-pumped the crucible for ten minutes before introducing nickelocene at a pressure of 5 ×10$^{-6}$ mbar through a leak valve (<1s deposition). Nickelocene tip functionalization was achieved by scanning an isolated nickelocene molecule in the constant current STM mode (set-point of $V_S$= 1.4 mV, $I$= 10 pA) with a Kolibri sensor.[47] Prior to conducting spectroscopy measurements, we tested the quality of the nickelocene tips on an Au(111) surface and only used stable nickelocene tips for the IETS measurements. The closest tip-sample distance set point (z=0) in IETS spectra presented in Fig. 4b,c was chosen to be $I$= 50 nA for bias voltage $V_S$= 25 mV (Fig. S7).

**Quantum chemical calculations.** The ground state optimized geometries of the free-standing and Au(111)-supported molecule **1** were obtained using the total energy spin-polarized DFT calculations with the AIMS code. The DFT calculations confirmed the absence of chemical bonding between molecule **1** and the metal surface. The NEVPT2/CASSCF(4,4) calculations were carried out in the def2-SVP basis using the Orca code.[48] In order to demonstrate that CAS(4,4) contains all strongly correlated orbitals, we have also performed CASSCF calculations with larger active spaces, namely CAS(8,8) and CAS(12,12) (the natural orbital occupation numbers are presented in SI). The high-level singlet-triplet excitation energy was computed by means of the DLPNO-NEVPT2 method with the default Orca settings and CASSCF(4,4) references. The CASSCF Dyson orbitals (also plotted in SI) and the singlet-triplet NTOs were computed using the in-house MOLMPS code,[49] interfaced to Orca. Unlike NTOs, we computed the Dyson orbitals with CAS(12,12) rather than CAS(4,4) due to symmetry-breaking problems in charged forms of molecule **1**.

**Acknowlagement**

J. Lu acknowledges the support from MOE grants (MOE2019-T2-2-044, MOE T2EP50121-0008, MOE-T2EP10221-0005) and Agency for Science, Technology and Research (A*STAR) under its AME IRG Grant (Project715 No. M21K2c0113). P. Jelinek acknowledge support from the CzechNanoLab Research Infrastructure supported by MEYS CR (LM2023051) and the GACR project no. 23-05486S. J.Wu acknowledges the financial support from A*STAR AME IRG grant (A20E5c0089) and NRF Investigatorship award (NRF-NRFI05-2019-0005). S. Song acknowledges the support from A*STAR under its AME YIRG Grant (M22K3c0094). This work was supported by the Czech Ministry of Education, Youth and Sports from the Large Infrastructures for Research, Experimental Development and Innovations project "IT4Innovations National Supercomputing Center-LM2015070", and the Computational Chemical Sciences Program of the U.S. Department of Energy, Office of Science, BES, Chemical Sciences, Geosciences and Biosciences Division in the Center for Scalable and Predictive methods for Excitations and Correlated phenomena (SPEC) at PNNL.

**Author Contributions**

J.Lu. supervised the project. S.S., J.W., and J.Lu. conceived and designed the experiments. L.V. and P.J. conceived the theoretical studies. S.S., H.Y., and M.T. performed the on-surface synthesis and LT-STM measurements. A.P.S., O.S. and P.J. performed the IETS measurements with $NiCp_2$ tip. G.L. and J.W. performed the organic synthesis of the precursor. A.M., L.V., D.S., M.K., J.B., and P.J. performed the theoretical calculations. J.Li. helped on the data presentation. S.S., P.J., and J.Lu. wrote the manuscript with input from all authors. All authors contributed to the scientific discussion.

**Competing interests**

The authors declare no competing interest.

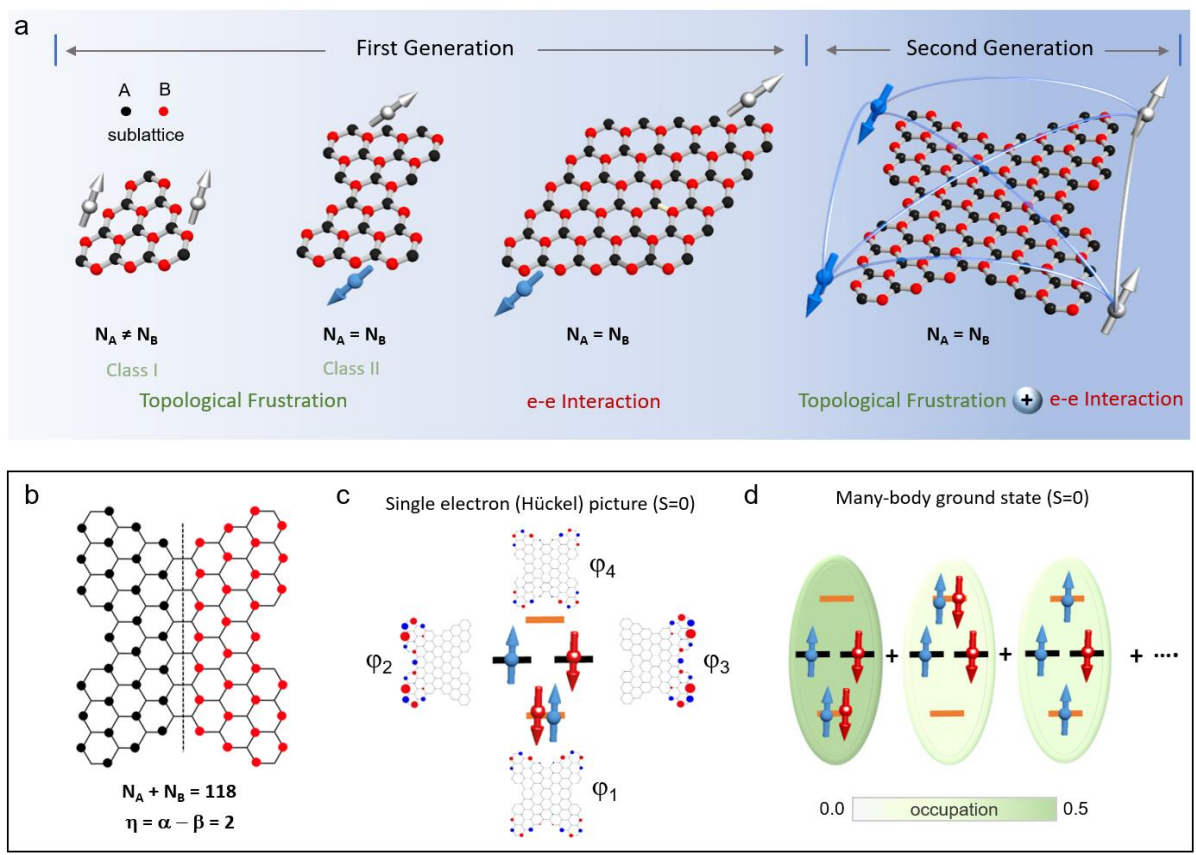

**Fig. 1 | Conceptual design principle of second-generation strongly-correlated polyradical nanographenes beyond a single π-magnetism origin a**. Schematic illustration of three representative first-generation open-shell nanographenes with different single π-magnetism origins including topological frustration class I (also termed as sublattice imbalance) and class II (both sublattices are topologically frustrated), and electron-electron interactions. Black and red dots represent the sublattices A and B, respectively, while $N_A$ and $N_B$ denote the number of sublattice A and B. **b**. Illustration of the 'nullity' of molecule **1**. For nanographenes with both sublattices topologically frustrated, such as Clar's goblet, the term 'nullity' ($\eta = \alpha - \beta$), where $\alpha$ and $\beta$ refer to the the maximum numbers of nonadjacent vertices and edges, respectively.[31] The sum of $\alpha$ and $\beta$ equals to N ($N_A + N_B$), the total number of carbon atoms in the nanographene. **c**. Hückel energy spectrum of **1** shows two zero-modes ($\varphi_2$ and $\varphi_3$) and two other states ($\varphi_1$ and $\varphi_4$). Hückel orbitals of **1** are presented at each energy level. The two states above and below the $E_F$ break spin symmetry due to the instability triggered by the electronic correlation. **d**. Many-body ground state energy diagram of the butterfly-shaped molecule **1**.

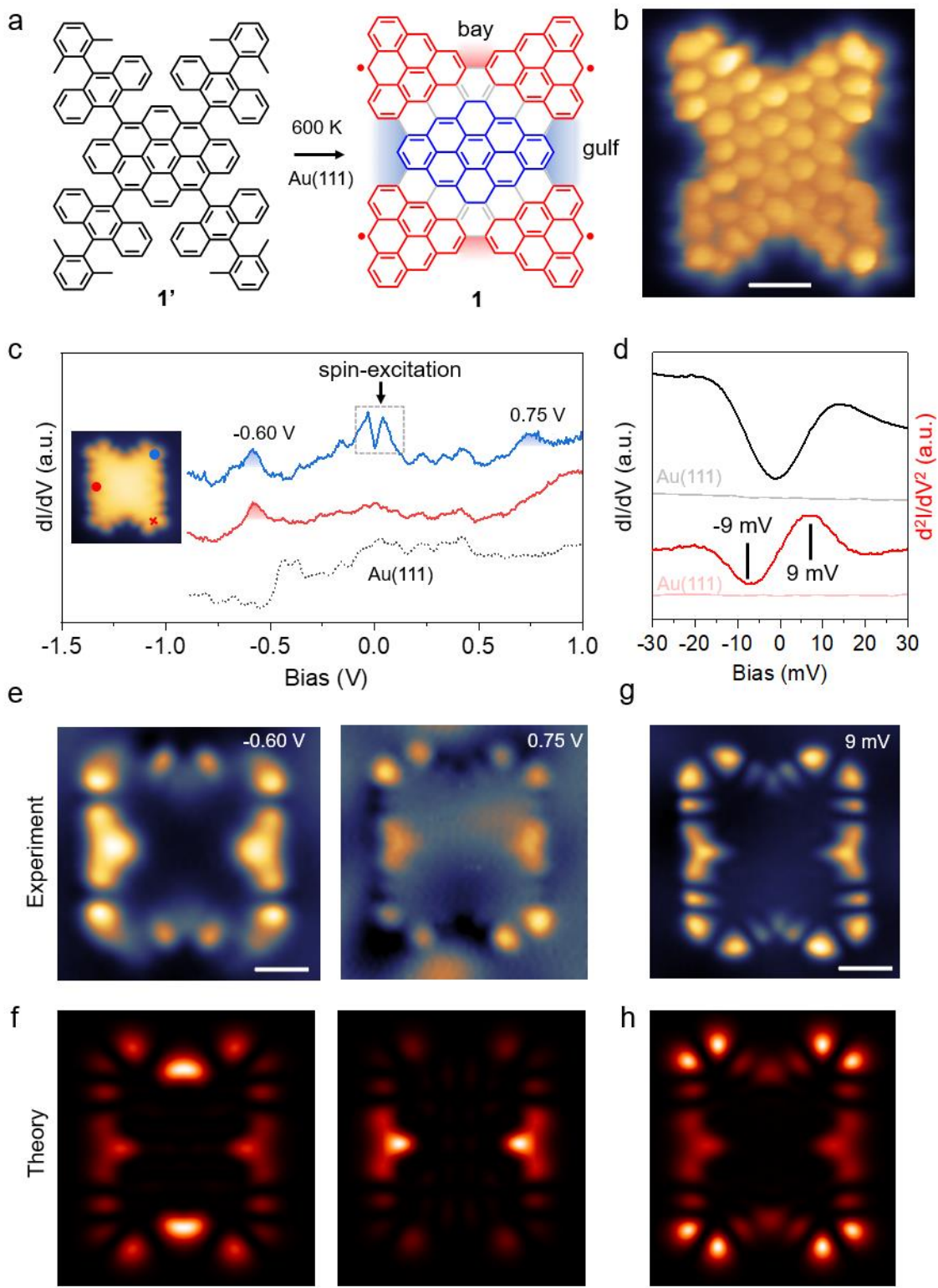

**Fig. 2 | On-surface synthesis and characterization of the 'butterfly' molecule 1. a.** Rational design of precursor **1′** for the on-surface synthesis of the product **1** on Au(111). **b.** Constant-height BRSTM image of molecule **1** taken with a CO-functionalized tip ($V_s$ = 20 mV). **c.** Point d$I$/d$V$ spectra of **1** were acquired at the position marked by the red and blue

dots in the inset of **c** using a metallic tip ($V_s$ = 1.5 V, $V_{ac}$ = 10 mV, and $f$ = 439 Hz). **d**. Point d$I$/d$V$ and IETS spectra of **1** taken at the position marked by the red cross in the inset of **c** using a metal tip ($V_s$ = 80 mV, $V_{ac}$ = 2 mV, and $f$ = 439 Hz). **e**. Constant current d$I$/d$V$ maps taken at -0.6 V and 0.75 V using a metallic tip ($I_t$ = 1 nA, $V_{ac}$ = 10 mV, and $f$ = 439 Hz). **f**. Calculated constant height d$I$/d$V$ maps at the conduction band and valence band of **1** from CASSCF Dyson orbitals. **g**. Spin-excitation map of **1** ($I_t$ = 1 nA, $V_{ac}$ = 2 mV, and $f$ = 439 Hz). **h**. Simulated spin-excitation maps of **1** by a sum of individual singlet-triplet CAS NTOs. Scale bars for **b**, **e**, and **g** are 0.5 nm.

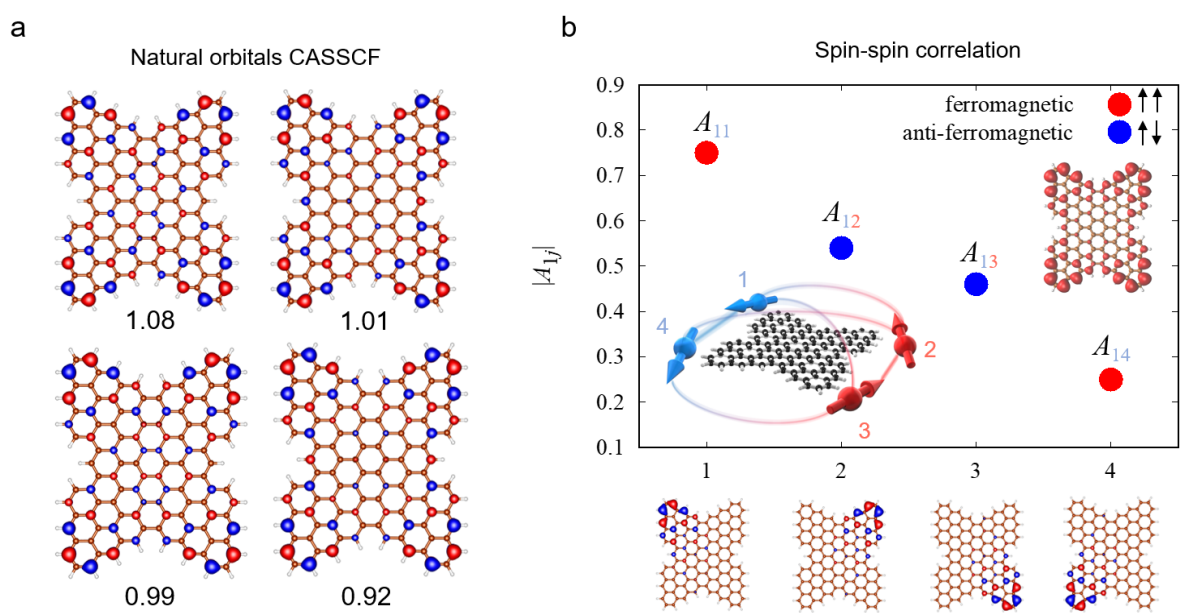

**Fig. 3| Spin correlation analysis of molecule 1 with the open-shell tetraradical character. a**. Four natural orbitals with different occupations in CAS(4,4). The number at the bottom of each natural orbital refers to the electron occupation **b**. Mutual coupling of individual unpaired spins in forming the singlet ground state of molecule **1** *via* theoretical analysis of the calculated spin-spin correlation function ($A_{ij} = \langle \hat{S}_i \hat{S}_j \rangle - \langle \hat{S}_i \rangle \langle \hat{S}_j \rangle$). Inset shows the conceptual illustrations of multiple spin interaction (left bottom). Inset (right top) presents the unpaired density distribution calculated by CASSCF(4,4). The four orbitals at the bottom of the panel represent the localization of the spins at each corner of molecule **1**.

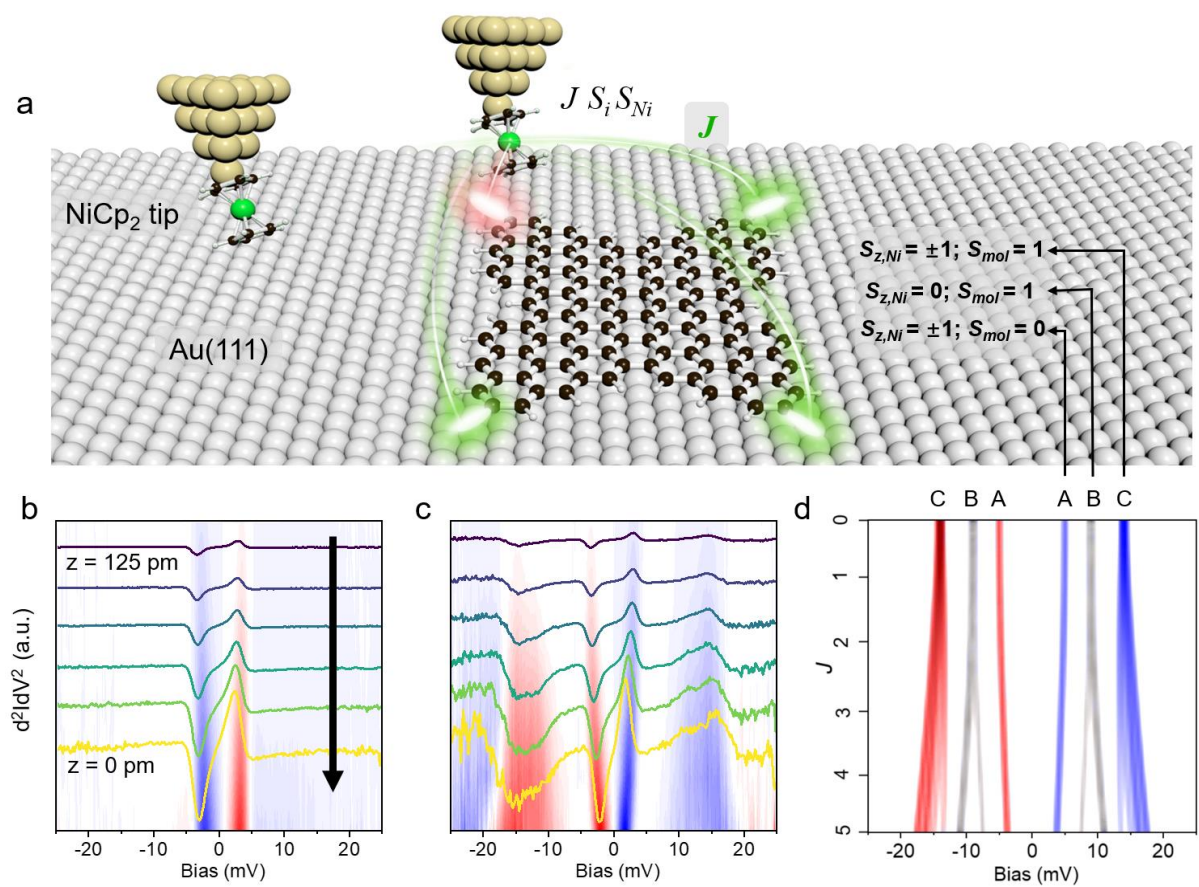

**Fig. 4 | Probe the π-magnetism of 'butterfly' 1 with a NiCp$_2$ functionalized tip. a.** Conceptual illustration of the measurement of molecule **1** with a NiCp$_2$ functionalized tip. **b.** IETS spectra plotted in a color scale taken over the bare Au(111) overlaid with the IETS spectra acquired at different tip-sample distances with a 25 pm decrease step. **c.** IETS spectra plotted in a color scale taken over the corner of 'butterfly' **1** overlaid with the corresponding IETS spectra taken at different tip-sample distances with the same 25 pm decrease step. IETS spectra lock-in parameters: $V_{ac}$ = 0.5 mV and $f$ = 723 Hz. **d.** Calculated IETS spectra plot of butterfly **1** as a function of the coupling strength $J$. States A and C (both red and blue) correspond to the experimental peak/dip signals shown in **c**, whereas state B (grey) is not visible.